# Control of Propagating Spin Waves via Spin Transfer Torque in a Metallic Bilayer Waveguide

Kyongmo An[1], Daniel R. Birt[1,2], Chi-Feng Pai[3], Kevin Olsson[1], Daniel C. Ralph[3,4], Robert A. Buhrman[3], and Xiaoqin Li[1,2*]

[1]*Department of Physics, University of Texas, Austin, TX 78712, USA*
[2] *Texas Material Institute, University of Texas, Austin, TX, 78712, USA*
[3] *Cornell University, Ithaca, New York, 14853, USA*
[4] *Kavli Institute at Cornell, Cornell University, Ithaca, New York, 14853, USA*

*\*Email: elaineli@physics.utexas.edu*

We investigate the effect of a direct current on propagating spin waves in a CoFeB/Ta bilayer structure. Using the micro-Brillouin light scattering technique, we observe that the spin wave damping and amplitude may be attenuated or amplified depending on the direction of the current and the applied magnetic field. Our work suggests an effective approach for electrically controlling the propagation of spin waves in a magnetic waveguide and may be useful in a number of applications such as phase locked nano-oscillators and hybrid information processing devices.

## I. INTRODUCTION

Research on propagating spin waves (SWs) or magnons in magnetic thin films has been invigorated in recent years because of their prominent role in several promising applications[1-3]. For example, SWs may serve as a link between phase locked nano-oscillators[4,5] or carry spin current in magnetic microstructures subject to a thermal gradient[6,7]. SWs have also been proposed for use in an information bus in a hybrid logic device[2] and for building logic gates[8]. A major obstacle to many of these applications is the fast damping or short propagation length of SWs in metallic magnetic layers. While parametric pumping can be used to amplify spin waves over a limited frequency range[9-11], a dedicated microwave source at a frequency twice of that of the SW is required. Alternatively, current induced magnetization manipulation via spin transfer torque (STT) has been proposed to compensate for the damping of SWs[12,13]. A high density of direct current (DC) that is spin polarized by passing through a ferromagnetic metallic nanowire has been proposed to compensate for SW damping via STT. However, the effect of SW amplification is small due to a small non-adiabatic STT, predicted and demonstrated to be less than 1%[13,14].

The search for more efficient STT materials has led to the investigation of metallic bilayer structures consisting of a magnetic layer and a nonmagnetic layer with strong spin-orbit coupling (e.g., Pt[15-18], Ta[19], and W[20]). In the nonmagnetic layer, spin polarized current is generated via the spin Hall effect (SHE), in which electrons with different spin states are deflected to opposite directions[21,22]. Such spin polarized current then exerts a STT on the adjacent magnetic layer, enabling control of magnetization dynamics via electric control. In fact, previous experiments have demonstrated that propagating SWs in an insulating yttrium iron garnet (YIG) film can be either amplified or attenuated via spin current injected from an adjacent Pt layer[17,23]. Due to experimental challenges imposed by the short SW propagation length in metals, such an effect has been elusive in metallic bilayers.

In this paper, we investigate the electric control of propagating SWs in a CoBFe/Ta bilayer waveguide. We use micro-Brillouin light scattering ($\mu$-BLS) to observe the amplitude of the SWs in the bilayer waveguide and its change due to an applied magnetic field and DC. We were able to achieve ~8% change in SW amplitude at a current density of ~$6 \times 10^6$ A/cm$^2$.

In bilayer structures, a number of factors contribute to the magnetization dynamics. In addition to the typical micro-magnetic contributions (e.g., dipole and exchange coupling, crystalline anisotropy) and the standard terms describing the coupling between magnetization and DCs (e.g., adiabatic and non-adiabatic STT), additional contributions have to be taken into account. Although there are still debates on the models best suited for describing the mechanism of the current-induced torque[24] (e.g., SHE in the heavy metal or a Rashba effect in the ferromagnetic layer[25]), these models often yield qualitatively similar results involving two torque components, respectively taking the form of $\boldsymbol{M} \times (\boldsymbol{j} \times \boldsymbol{n})$ and $\boldsymbol{M} \times [\boldsymbol{M} \times (\boldsymbol{j} \times \boldsymbol{n})]$, where $\boldsymbol{M}$ is the magnetization; $\boldsymbol{j}$ is the in-plane current density, and $\boldsymbol{n}$ is a unit vector along the interface normal toward the magnetic layer. The first term has the same form as the precessional torque around an effective field along the direction of $-(\boldsymbol{j} \times \boldsymbol{n})$, and is thus called a field-like torque. The second term can act to either increase or decrease damping, and is thus called a damping-like torque. While a rich variety of physical phenomena exist in the bilayer structure subject to DC, the key observation of our experiments (i.e. change of SW amplitude) can be explained by the damping-like torque arising from the SHE.

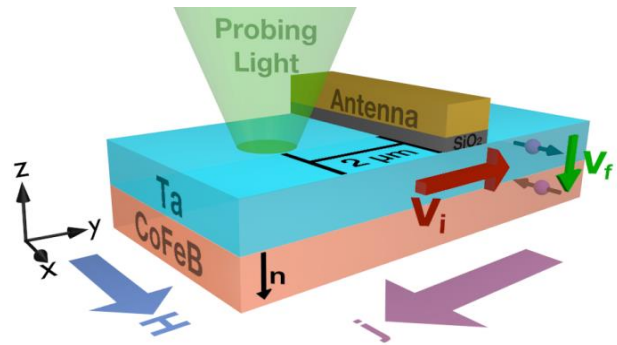

Fig. 1: Schematics of spin polarized current generation in a bilayer and the experimental setup of micro-BLS.

We first describe the basic concepts and physical processes involved in our experiments. As illustrated in Fig. 1, a magnetic (CoFeB) waveguide is covered by a nonmagnetic layer (β-Ta), and a DC passes through both layers. An external magnetic field is applied in the plane of the layered structure and perpendicular to the waveguide along the $x$ direction. In this configuration, a microwave antenna on top of the nonmagnetic layer excites Damon-Eshbach surface SW modes that propagate along the waveguide[26]. For a DC $j$ flowing along the $-y$ direction, the electrons move along the $y$ direction. The spin polarized electrons deflected toward magnetic layer along the $-z$ direction possess magnetic moment $\sigma$ determined by $\sigma \parallel -\theta_{SH} v_i \times v_f$ [21] where $v_i$ is the initial electron velocity driven by the DC, $v_f$ is the deflected electron velocity via SHE, and $\theta_{SH}$ is the spin Hall angle of Ta, which is defined as the ratio between the spin and charge current density, i.e., $\theta_{SH} = j_s/j_e$. In our experiments, $v_i \parallel -j$ and $v_f \parallel n$, thus we can rewrite $\sigma \parallel \theta_{SH} j \times n$. The direction of $\sigma$ is then along $-x$ direction because $\theta_{SH} < 0$ in Ta[19]. The direction of the damping-like torque exerted by the polarized electrons can be rewritten as $\tau \parallel (M \times \sigma) \times M$, which tilts the $M$ away from the external field for $H \parallel x$ (i.e., $H > 0$) and $j \parallel -y$ (i.e., $j < 0$), thus leading to reduced damping and amplified SWs. Either reversing the direction of $H$ or $j$ leads to attenuation of the SW in this simple picture.

## II. BILAYER STRUCTURE AND EXPERIMENTAL TECHNIQUE

Specifically, our sample consists of $Co_{40}Fe_{40}B_{20}(10)$/ Ta(10) sputter-deposited onto a thermally oxidized Si substrate in the same chamber used in Ref. 19. The numbers in parentheses represent layer thickness in nanometers. Following the deposition, the bilayer structure was patterned into 8 μm-wide and 270 μm-long waveguide. A 6 μm-wide Cu(150)/Au(10) antenna separated by an 80nm-thick $SiO_2$ layer was then created on top of the bilayer waveguide. Based on the geometry of the bilayer structure and the measured resistance, the resistivity equals 175 ± 6 μΩ cm, close to the average resistivity of β-Ta (~190 μΩ cm) and CoFeB (~170 μΩ cm)[19].

We briefly describe our experimental approach. The μ-BLS technique offers high spatial resolution, sensitivity, and dynamical range as demonstrated in previous experiments[27-29]. The BLS signal arises from the inelastic scattering of light by SWs, and is proportional to the intensity of the SWs, i.e., proportional to the SW amplitude squared. A linearly polarized, single frequency laser beam at 532 nm was focused to a spot size of ~ 1 μm in diameter on the sample. The laser power was approximately 1 mW. The scattered light with orthogonal polarization was collected and sent to a Sandercock multi-pass tandem interferometer to resolve the inelastic scattering from SWs.

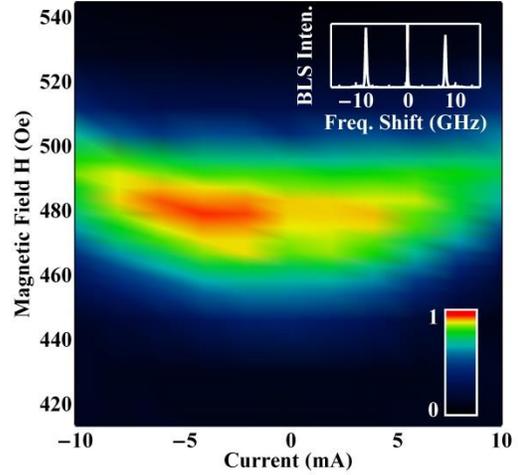

Fig. 2: SW amplitude as a function of DC and external magnetic field. Inset: A raw BLS spectrum showing both the Stokes and anti-Stokes peaks scattering from magnons.

An imaging-based position correction algorithm was implemented to avoid beam drift during data acquisition.

## III. EXPERIMENTAL RESULTS AND DISCUSSION

A representative BLS spectrum for SWs excited via the antenna with a driving frequency of 8 GHz and a microwave power of 2 mW is shown in the inset of Fig. 2, where both the Stokes and anti-Stokes peaks are observed. The linewidth of the raw BLS spectrum does not provide any information about damping, because it is limited by the frequency resolution of our interferometer (~ 0.8 GHz).

For the fixed excitation frequency of 8 GHz and the chosen spatial location (shown in Fig. 1), we scanned the applied magnetic field and DC passing through the bilayer waveguide to provide a two-dimensional (2D) map, as shown in Fig. 2. The color corresponds to the square root of the integrated BLS signal from the Stokes peak. There are two prominent features in this 2D field-current map. First, the H field corresponding to the maximal SW amplitude shifts to higher values when a DC is applied due to heating. Secondly, the amplitude of the SW is modified by the DC, but not in a monotonic way. A key difference between our experiments and those previously performed on bilayer structures based on magnetic insulators[17, 23] is that the Joule heating effect is pronounced in metallic bilayer structures, which we account for in our data analysis.

We take line cuts in the 2D map to examine the field and DC dependence of SW amplitude in detail. We first plot the SW amplitude as a function of the applied magnetic field at 10 and -10 mA in Fig. 3a. We discuss two features in the field dependence. First, there is a clear difference in the SW amplitudes when DCs of opposite sign are applied, which is the key evidence for DC-controlled SW amplification and attenuation. The direction of DC for the SW amplification is consistent with that expected from the

SHE, as illustrated in Fig. 1. Second, the DC effect on the magnetic dynamics is different for magnetic fields below and above the maximum of the BLS signal peak, with a more pronounced effect at lower magnetic fields where propagating SWs are excited.

We then fit the field dependent spectra using eq. (S1) in Ref. 30 and extracted the effective magnetization $4\pi M_{eff}$ at each applied DC. A spin wave mode that corresponds to the lowest transversely quantized mode ($k_x = \frac{\pi}{w}$) and the longitudinal uniform procession mode ($k_y = 0$) is described by[30]

$$f = \frac{\gamma}{2\pi}\sqrt{H_R\left[H_R + 4\pi M_{eff}\left\{1 - \frac{1}{2}\left(\frac{\pi}{w}\right)d\right\}\right]} \quad (1)$$

where $\gamma$ is the gyromagnetic ratio. $w$ and $d$ are the width and thickness of the CoFeB layer. Using the extracted $4\pi M_{eff}$ and Eq. (1), we extracted $H_R$ as a function of the DCs shown in Fig. 3b. We note that $H_R$ is approximately the magnetic field that corresponds to the peak of the BLS signal in magnetic field dependent spectra (e.g. Fig. 3a)[30]. As the DC increases, there is a clear shift in $H_R$, consistent with the heating effect. Joule heating raises the temperature of the magnetic waveguide, which leads to a reduction of the effective magnetization $M_{eff}$, thus an increase in $H_R$ as required by eq. (1). Based on the shift of $13 \pm 3$ Oe in $H_R$ and the corresponding change of $M_{eff}$, a temperature change of $41 \pm 3$ K was calculated with the Bloch $T^{3/2}$ law[31]. This value is close to the temperature change $\Delta T = 42.5 \pm 6.1$ K calculated from temperature dependent resistance using $\beta\Delta T = (R_{10mA} - R_{0mA})/R_{0mA}$, where $R_{10mA}$ = 2978 Ω, $R_{0mA}$ = 2960 Ω and $\beta$ is the measured temperature coefficient of resistivity of the bilayer structure, $(145 \pm 21) \times 10^{-6}$/K.

To isolate the change in the SW amplitude due to the STT arising from the SHE, we remove the effect of heating by comparing the SW amplitudes at two DCs of the same magnitude but opposite directions at a particular magnetic field. The following analysis assumes that the BLS signal arises mainly from a single transversely quantized SW mode ($f = 8\ GHz$) with a wave vector determined by the dispersion relation described in detail in Ref. 30. The SW amplitudes at positive and negative DCs can be written as a sum of several contributions

$$\begin{aligned} A_+ &= A_0 + \Delta A_{Heat} - \Delta A_{SHE} \\ A_- &= A_0 + \Delta A_{Heat} + \Delta A_{SHE} \end{aligned} \quad (2)$$

where $A_0$ is the SW amplitude at 0 mA and $A_-(A_+)$ corresponds to the SW amplitude for $j < 0$ ($j > 0$). $\Delta A_{Heat}$ and $\Delta A_{SHE}$ are the contributions to the SW amplitudes due to the heating and SHE, respectively. We also use the fact that $\Delta A_{Heat}$ and $\Delta A_{SHE}$ are an even function and odd function of DC[15,32], respectively over the range of current density investigated here. The effect of heating is effectively removed by subtracting $A_+$ from $A_-$.

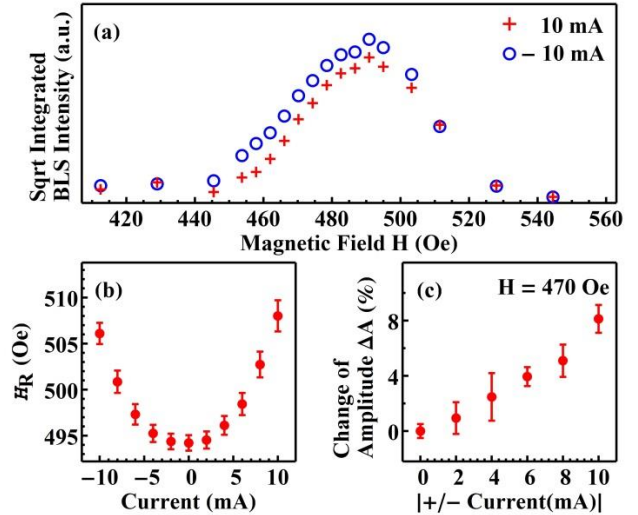

Fig. 3: Detailed analysis of data presented in Fig. 2. (a) Magnetic field dependent SW amplitude when DCs of opposite directions were applied. (b) Extracted $H_R$ as a function of DCs. (c) Change in the propagating SW amplitude at a particular magnetic field as a function of DCs obtained by subtracting SW amplitudes for DCs of opposite direction.

By normalizing with the sum of these two quantities, we obtain the fractional change of SW amplitudes due to the SHE at increased temperatures.

$$\Delta A = \frac{A_- - A_+}{A_- + A_+} = \frac{\Delta A_{SHE}}{A_0 + \Delta A_{Heat}} \quad (3)$$

The percentage change in SW amplitude at $H$ = 470 Oe from $\Delta A$ is plotted in Fig. 3c. An approximate linear dependence was observed up to the maximal current density applied. Because $\Delta A_{Heat} < 0$, the apparent SW amplitude change in Fig. 2 is not monotonic as a function of DCs. Only when the heating effect is removed by subtracting $A_+$ from $A_-$, can the SHE be revealed. We note that other effects such as the Oersted field and wavelength dependent STT efficiency[15] may lead to an asymmetric shift with respect to DCs in the field dependent spectra, thus affecting the SW amplitude at a particular DC. However, the change in SW amplitude due to the net spectral shift is negligible in our devices, given the small size of the asymmetric shift in $H_R$ as shown in Fig. 3b compared to that observed in a control sample[30].

We now estimate the change in damping following the approach outlined in Ref. 17. The spin wave propagating with amplitude $u(x,t)$ can be modeled as:

$$i\frac{\partial u}{\partial t} = \left[\omega_0 + v_g\left(-i\frac{\partial}{\partial x} - k_0\right)\right]u - i\eta u - i\Delta\eta u \quad (4)$$

where $\omega_0$, $k_0$, and $v_g$ are the angular frequency, wave vector, and group velocity, respectively. $\Delta\eta$ represents the STT induced change in the SW decay rate $\eta$. Only the lowest transversely quantized mode was considered in this analysis. We show that the extracted $\Delta\eta$ changes by ~2% if other SW modes are considered[30]. In our notation, one has $\Delta\eta < 0$ ($\Delta\eta > 0$) for $j < 0$ ($j > 0$). The SW amplitude at $x = L$ when a DC is applied can be written as:

$$A_{\mp} = A_0 e^{\pm\frac{|\Delta\eta|L}{v_g}} \quad (5)$$

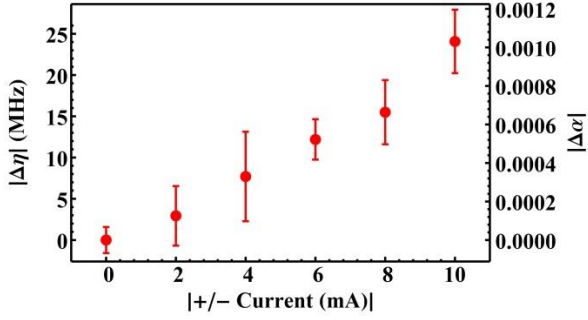

Fig. 4: The change in decay rate (left axis) and damping constant (right axis) as a function of DC. The error bars quoted are for $\Delta\alpha$ and mainly arise from uncertainty in $\Delta A$.

When we compare the cases with equal DCs of opposite directions, we obtain:

$$\Delta A = \tanh\frac{|\Delta\eta|L}{v_g} \quad (6)$$

From the data presented in Fig. 3c, we extracted $\Delta A$ at different DCs and calculated $|\Delta\eta|$ as shown in Fig. 4. In this calculation, we have used $L = 2\ \mu m$ corresponding to the distance from the edge of the antenna to the detection point. The group velocity corresponding to the SW excited at 8 GHz was calculated from the dispersion curve[30]. One can calculate the change in damping from $\Delta\eta$ using $\Delta\alpha\gamma(H + 2\pi M_{eff}) = \Delta\eta$.[9] The calculated $|\Delta\alpha|$ is also plotted in Fig. 4, where we used $\gamma/(2\pi) = 2.92$ GHz/kOe,[33] $H = 470$ Oe, and $4\pi M_{eff}$ obtained from the fitting procedure described in Ref. 30. If we use an approximate value of $\alpha = 0.008$ from the literature[19, 33], the maximal change in damping is about 13% at $I = 10$ mA, corresponding to the current density of $6.3 \times 10^6$ A/cm$^2$. In principle, higher driving current can be used to achieve further compensation over damping. The maximal current density of $\sim 6 \times 10^6$ A/cm$^2$ is limited by the damage threshold for our current device. In comparison, in a previous experiment on controlling thermal magnetic fluctuations in a Py/Pt micro-disk via STT, the damping can be modified by a factor of 4 at a current density close to $10^8$ A/cm$^2$.[32] In fact, auto-oscillations due to a complete compensation of damping have been observed when localized SWs were excited[34, 35]. This difference in damping control may arise from the different requirements in controlling the localized SWs and propagating SWs, different interface properties, as well as the thicker ferromagnetic layer, the wide width of the waveguide, and the low damage threshold for our current devices.

Finally, we investigate the BLS signal dependence on microwave power applied to the antenna when a DC passes through the waveguide in Fig. 5 (data taken from a different device). We plot the result for $I = -8$ mA vs. $I = 8$ mA for $H = 446$ Oe in Fig. 5a. The arrows indicate the power of 2 mW chosen in the previous sets of measurements presented in the paper, which lies in the low power regime where BLS intensity increases linearly with microwave power. The SW amplitude at $I = -8$ mA was higher over the whole power range, and the values of $|\Delta A|$ showed small variations averaging ~7%. When the magnetic field $H$ was reversed, the SW amplitude at $I = 8$ mA was higher (Fig. 5b), consistent with the damping-like STT determined by $(M \times \sigma) \times M$ (illustrated in the insets of Fig. 5). Thus, one may control the SW amplitude by changing either the direction of $H$ or DC. The same microwave power dependence measurement was performed on a control sample consisting of CoFeB(10)/Ta(1), where the top Ta(1) layer is oxidized in air and mainly served as a protection layer. There was no detectable difference in the BLS signal between DCs of 8 mA and $-8$ mA (Fig. 5c). Effects such as the presence of multiple SW modes and the temperature dependent SW wavelength are also present in the control sample. The marked difference observed in the control sample suggests that the spin polarized current generated in the Ta(10) layer was the key to the SW amplitude and damping control observed in the sample of interest.

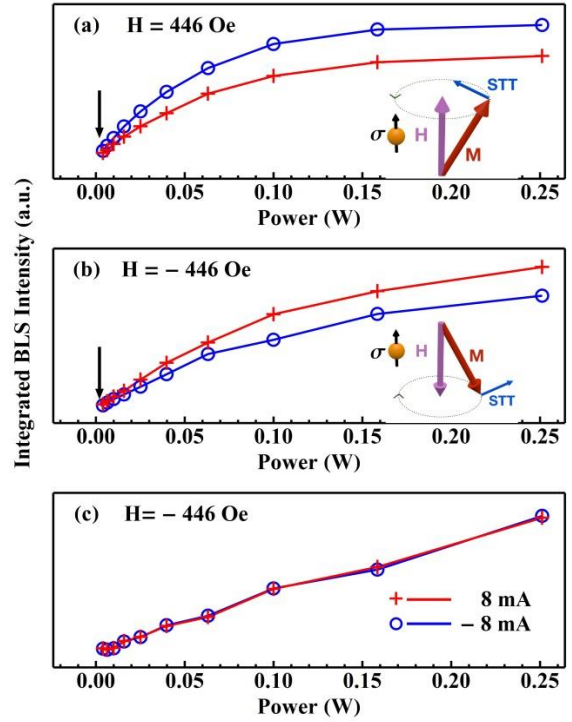

Fig. 5: BLS intensity as a function of the microwave power applied across the antenna for 8 mA (red crosses) and -8 mA (blue open circles) for (a) $H > 0$, (b) $H < 0$, and (c) $H < 0$ from a CoFeB(10)/Ta(1) waveguide. The arrows refer to the microwave power used in other experiments presented in this paper. The insets illustrate STT arising from the spin polarized electrons $\boldsymbol{\sigma}$ with $j > 0$ for (a) $H > 0$ and (b) $H < 0$, respectively.

## IV. CONCLUSION

In conclusion, we have studied the effect of DCs on propagating SWs in a CoBFe/Ta bilayer waveguide using $\mu$-BLS technique. After removing the effect of heating in the analysis, we observed an 8% change of SW amplitude at a reasonable DC density in a device that has not been fully optimized. Further improvement in device design and fabrication may eventually lead to complete SW damping compensation[36], thus opening many exciting opportunities in spintronics and magnonics.

## ACKNOWLEDGMENTS

We would like to thank Mark Stiles for suggesting the experiment and Maxim Tsoi for helpful discussions. We gratefully acknowledge financial support from the following sources: AFOSR FA9550-08-1-0463 and AFOSR FA-9550-08-1-0058. DB acknowledges a fellowship from the NSF-IGERT program via grant DGE-0549417.